\documentclass[aps,prb,twocolumn,floatfix,showpacs,superscriptaddress]{revtex4-2}
\usepackage{graphics}
\usepackage{epsfig}
\usepackage{times}
\usepackage{bm}
\usepackage{braket}
\usepackage{color,float}
\usepackage{amsmath}
\usepackage{amssymb}
\usepackage[caption = false]{subfig}
\usepackage{color,graphicx,pstricks}
\usepackage{bbm}
\usepackage{array}

\renewcommand{\i}{{\rm i}}

\usepackage{amsmath}	
\begin{document}

\title{Disorder effects in the Kitaev-Heisenberg model}

\author{Ayushi Singhania} 
\affiliation{Institute for Theoretical Solid State Physics, IFW Dresden, 01069 Dresden, Germany}

\author{Jeroen van den Brink}
\affiliation{Institute for Theoretical Solid State Physics, IFW Dresden, 01069 Dresden, Germany}
\affiliation{Department of Physics, Technical University Dresden, 01069 Dresden, Germany}

\author{Satoshi Nishimoto} 
\affiliation{Institute for Theoretical Solid State Physics, IFW Dresden, 01069 Dresden, Germany}
\affiliation{Department of Physics, Technical University Dresden, 01069 Dresden, Germany}

\begin{abstract}
We study the interplay of disorder and Heisenberg interactions in Kitaev model on honeycomb lattice. The effect of disorder on the transition between Kitaev spin liquid and magnetic ordered states as well as the stability of magnetic ordering is investigated. Using Lanczos exact diagonalization we discuss the consequences of two types of disorder: (i) random-coupling disorder and (ii) singular-coupling disorder. They exhibit qualitatively similar effects in the pure Kitaev-Heisenberg model without long-range interactions. The range of spin liquid phases is reduced and the transition to magnetic ordered phases becomes more crossover-like. Furthermore, the long-range zigzag and stripy orderings in the clean system are replaced by their three domains with different ordering direction. Especially in the crossover range the coexistence of magnetically ordered and Kitaev spin-liquid domains is possible. With increasing the disorder strength the area of domains becomes smaller and the system goes into a spin-glass state. However, the disorder effect is different in magnetically ordered phases caused by long-range interactions. The stability of such magnetic ordering is diminished by singular-coupling disorder, and accordingly, the range of spin-liquid regime is extended. This mechanism may be relevant to materials like $\alpha$-RuCl$_3$ and H$_3$LiIr$_2$O$_6$ where the zigzag ground state is stabilized by weak long-range interactions. We also find that the flux gap closes at a critical disorder strength and vortices appears in the flux arrangement. Interestingly, the vortices tend to form kinds of commensurate ordering.
\end{abstract}
	
\date{\today}
	
\maketitle


\section{Introduction}
Quantum spin liquids (QSLs) are entangled states of matter, where quantum fluctuations prevent the formation of magnetic order down to lowest temperatures. They have attracted substantial attention in the developments of topological quantum computing due to its remarkable properties, including long range entanglement, topological ground state and fractionalized excitations~\cite{balents2010spin,takagi2019concept}. 
One of the most promising theoretical models to realise QSLs, was proposed by Kitaev in his seminal paper~\cite{kitaev2006anyons}. The model is based on  a two dimensional honeycomb lattice with bond anisotropic interactions among quantum spin-1/2 particles. Remarkably, such Kitaev-type interactions are predicted to be realized in transition metal oxides with spin-orbit couplings~\cite{jackeli2009mott}. Since then, layered honeycomb materials A$_2$IrO$_3$ (A = Na, Li)~\cite{singh2012relevance,chaloupka2010kitaev} and $\alpha-$RuCl$_3$  \cite{koitzsch2016j} have been thoroughly investigated as the candidate Kitaev materials~\cite{winter2017models,trebst2017kitaev}. Even though thermodynamic and dynamical properties of these materials match the theoretical prediction of Kitaev model, all the candidate materials are known to magnetically order at low temperatures \cite{choi2012spin,williams2016incommensurate}. Magnetic orders in such materials are manifested by the interactions beyond the  Kitaev interactions \cite{rau2014generic,hwan2015direct,sears2015magnetic}. 



In addition, presence of disorder like vacancies, impurities, lattice distortions, and stacking faults, inevitable in real materials, are often responsible for the instability of QSL \cite{freitas2022gapless, andrade2020susceptibility,kao2021disorder}. Recently, a new class of intercalated compounds have been synthesized which are even more susceptible to disorder H$_3$LiIr$_2$O$_6$, Cu$_2$IrO$_3$, and Ag$_3$LiIr$_2$O$_6$ \cite{bahrami2021effect, bahrami2022metastable}. Role of disorder in Kitaev materials has gained special interest since the experimental observation of low temperature divergence in specific heat of the proximate Kitaev material, H$_3$LiIr$_2$O$_6$, is understood as the consequence of bond disorder in the Kitaev model \cite{kitagawa2018spin, knolle2019bond}. Theoretically, it is known that bond randomness in the Kitaev model affects the low-energy dynamical spin-correlators~\cite{zschocke2015physical}. Thermodynamic properties like specific heat and susceptibility, spin transport, and spin dynamics at low energy excitations of the Kitaev model with bond randomness and site dilution have been also studied at finite but low temperatures~\cite{nasu2021spin,nasu2020thermodynamic,nasu2020thermodynamic,andrade2020susceptibility,willans2010disorder,willans2011site, kao2021vacancy}. 
Moreover, the effect of bond randomness on the zero-temperature topological properties of the Kitaev model~\cite{yamada2020anderson,dantas2022disorder} as well as the spin excitations in a diluted Kitaev system containing spin vacancies and bond randomness~\cite{do2020randomly} have been discussed. 

While a lot of studies have been devoted to the thermodynamics (relevant for experiments) of disordered Kitaev models, very little is understood about the ground state properties of the Kitaev model in the presence of disorder. Furthermore, presence of any type of disorder such as impurity or stacking faults must affect the interactions beyond the Kitaev limit. However, even less is known about the disorder effects, for example, in the presence of Heisenberg interaction. Once the Heisenberg interaction is added in the Kitaev model (referred as Kitaev-Heisenberg model), variety of magnetically ordered phases are stabilized as a consequence of order by disorder mechanism. Understanding the effect of disorder in the Kitaev-Heisenberg model is crucial from the point of view of interplay between frustration and disorder.

Motivated by this, we employ numerical diagonalization of 24-site cluster on a honeycomb lattice, to study the effects of disorder in the ground state of Kitaev-Heisenberg model. This serves as a putative minimal model for several QSL candidates. We emphasise on two different types of bond-disorder introduced in both Kitaev $and$ Heisenberg interactions. In the pristine Kitaev-Heisenberg model, Kitaev QSL state survives the small perturbations like Heisenberg interaction\cite{chaloupka2013zigzag}. Here, we show that the QSL remains robust against weak disorder and Heisenberg interaction, the stability of QSL region is narrowed with increasing disorder strength. The long-range zigzag and stripy orderings are replaced by a mixture of their three domains with different ordering direction. The area of domains is reduced with increasing disorder strength and the system goes into a spin-glass state. Moreover, we show that a flux state can be induced by bond-disorder in the ground state. The transition from flux-free state to random flux state in the presence of bond disorder has already been predicted just in the Kitaev limit~\cite{zschocke2015physical, nasu2021spin}. Interestingly, this random-flux state persists even for a small admixing of nearest-neighbor Heisenberg interaction. We also find that disorder in presence of further neighbor interaction can indeed result in a QSL or disordered state. This is relevant to experimental observations of H$_3$LiIr$_2$O$_6$~~\cite{kimchi2018scaling}.


The paper is organized as follows: In Sec. II, we begin by introducing the type of disorders considered in the Kitaev-Heisenberg model and the details of numerical methods applied. In Sec. III, we present the results of disorder effects on ground state phases in K-H parameter space. Moreover, we also discuss the effect of further neighbour interactions on the ground state of disordered KH model. Section IV is devoted to the summary.


\begin{figure}[tbh]
\includegraphics[width=0.9\columnwidth]{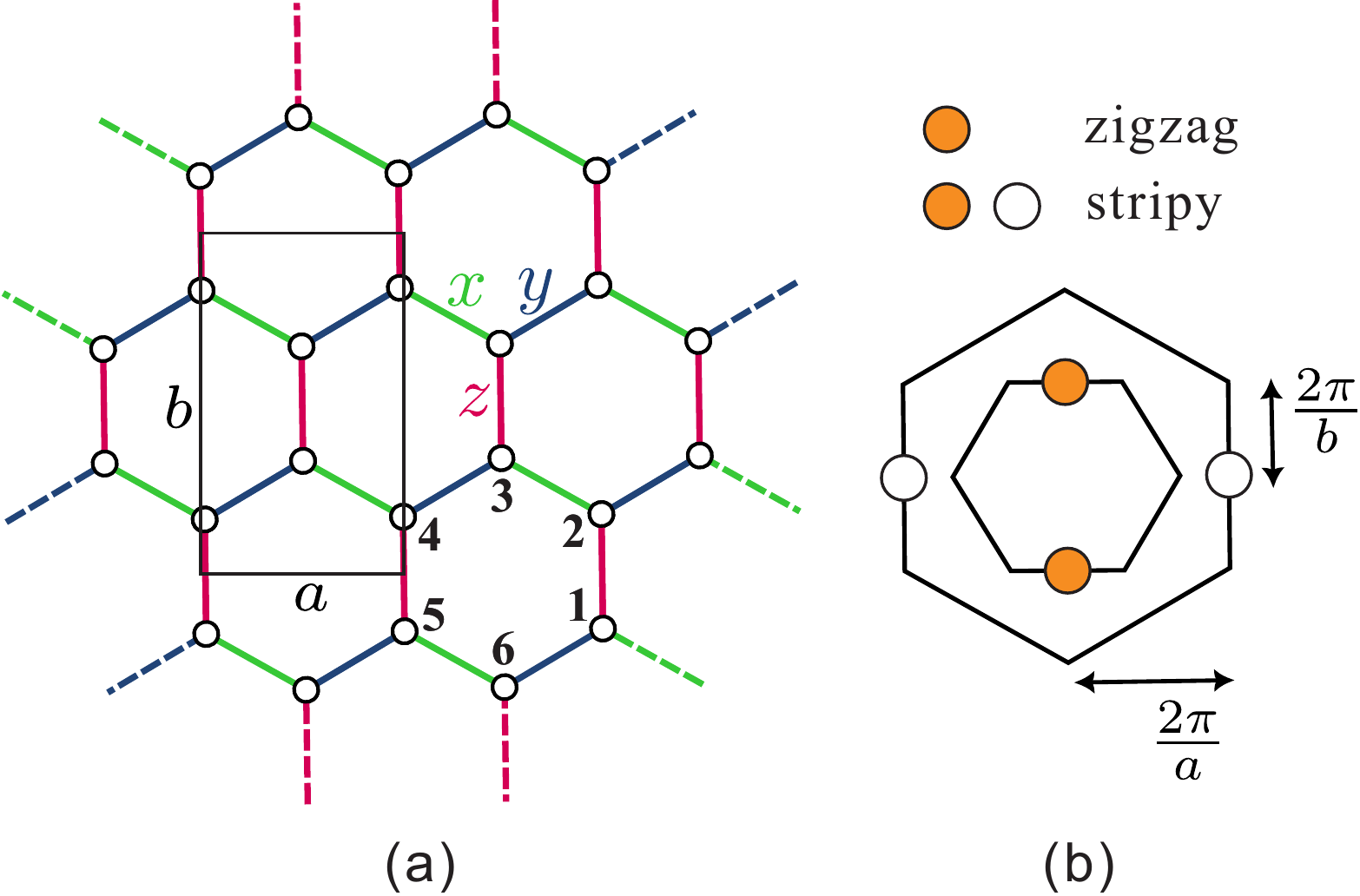}
	\caption{(a) 24-site periodic cluster used in our exact-diagonalization calculations. A black rectangle shows the unit cell. Numbers 1-6 correspond to labeling for the hexagonal
	plaquette operator [Eq.~\eqref{fluxop}].
	(b) Corresponding Brillouin zone, where inner hexagon shows first Brillouin zone. Magnetic Bragg peaks for zigzag and stripy states are also shown.}
	\label{fig:model}
\end{figure}

\section{Model and method}

\subsection{Kitaev-Heisenberg Model}

The Kitaev honeycomb model is described by anisotropic Ising-type interactions among spin-1/2 degrees of freedom on a honeycomb lattice, where the interaction in spin space ($\gamma=x,y,z$) is dictated by the bond direction. The bond-directional Ising-type interactions are called Kitaev interactions. The model is exactly solvable with the ground state being QSL and its elementary excitations described by Majorana fermions coupled to the $Z_2$ gauge field. In general, the systems exhibiting the Kitaev interactions are referred as Kitaev materials. However, as revealed from extensive research, real Kitaev materials possess  interactions beyond Kitaev, such as diagonal exchange couplings, i.e., Heisenberg terms, as well as off-diagonal couplings, i.e., so-called $\Gamma$ terms \cite{rau2014generic,hwan2015direct}. Nevertheless, a QSL ground state state is maintained even in the presence of weak Heisenberg terms~\cite{chaloupka2013zigzag}. In order to derive a common understanding on the disorder effects in Kitaev systems, we here restrict ourselves to models with the Heisenberg and Kitaev terms, i.e., the Kitaev-Heisenberg model.

\subsubsection{Clean case}

The Hamiltonian of the Kitaev-Heisenberg model reads as
\begin{equation}
\mathcal{H}_{\rm KH}=\sum_{\langle ij \rangle\gamma} 2K_{ij}^\gamma S_i^\gamma S_j^\gamma +  \sum_{\langle i,j \rangle} J_{ij} \mathbf{S}_i \cdot \mathbf{S}_j
\label{model}
\end{equation}
where $K_{ij}^\gamma$ is the Kitaev interaction of nearest neighbor spins on three different bonds $\gamma=x,y,z$ and $J_{ij}$ is the Heisenberg interaction between nearest neighbours. We assume all the Kitaev couplings are isotropic with $K_{ij}^x=K_{ij}^y=K_{ij}^z=K$ and the nearest-neighbor Heisenberg coupling is constant with $J_{ij}=J$ in the clean case. For simplicity, we parameterize as following: $J=A\cos\phi$ and $K=A\sin\phi$, where $A=\sqrt{J^2+K^2}$ is used as an energy unit. The ground state phase diagram is known to host four ordered and two QSL phases as a function of $\phi$: N\'eel ($-0.19\pi\lesssim\phi\lesssim0.49\pi$), AFM QSL ($0.49\pi\lesssim\phi\lesssim0.51\pi$), zigzag ($0.51\pi\lesssim\phi\lesssim0.90\pi$), FM ($0.90\pi\lesssim\phi\lesssim1.40\pi$), FM QSL ($1.40\pi\lesssim\phi\lesssim1.58\pi$), and stripy ($1.58\pi\lesssim\phi\lesssim1.89\pi$) ~\cite{chaloupka2013zigzag}. In this paper, we consider two types of bond disorders explained below, and mainly focus on the effect of disorders around the two Kitaev QSL phases, i.e., around $\phi = \pi/2, 3\pi/2 $.

\subsubsection{Random-coupling disorder}

The first type of disorder is a random distribution of bond strength (called ``random-coupling disorder''), which commonly appear due to the structural disorders in crystals. This can be achieved by changing the parameter $A \rightarrow (A+\delta A_{ij}$) for all nearest-neighbor bonds, such that $J \rightarrow (A+\delta A_{ij}) \cos \phi$ and $K \rightarrow (A+\delta A_{ij}) \sin \phi$. Namely, the Kitaev and Heisenberg couplings of 36 nearest-neighbor bonds in our 24-site periodic cluster (see Fig~\ref{fig:model}(a)) take different values from each other. Nevertheless, note that the ratio between $K$ and $J$ remains as $K/J=\tan\phi$ for each bond. The random deviation from the original bond, $\delta A_{ij}$, is defined by a Gaussian distribution $N(0,\sigma^2)$ where $\sigma$ is the standard deviation. The disorder strength is controlled by $\Delta=2\sigma$. The cases of $\Delta=0.125$, $0.25$, and $0.5$ are studied in this paper. For this kind of bond disorder, a box probability has been also frequently used in the previous studies but just for the Kitaev limit, $K/J\to\infty$. Although the effect of disorder with a box probability may be somewhat a little more drastic than that with a Gaussian distribution, their results are expected to be qualitatively similar.

\subsubsection{Singular-coupling disorder}

The second type of disorder is a random mixture of two kinds of bond strengths (called ``singular-coupling disorder''). This can be achieved by replacing a part of the original bonds with $r$-times stronger bonds, where the fraction of replaced bonds to the total bonds is defined by $n$. In other words, randomly selected fraction $n$ of original bonds are changed as $J \rightarrow rJ$ and $K \rightarrow rK$. We call $n$ disorder density and $r$ disorder strength. Since we use a finite-size periodic cluster as described below, it would be reasonable to keep equal number of replaced $x$, $y$ and $z$ bonds. In fact, we vary the number of replaced bonds in steps of 3. The case of $n=0$ and/or $r=1$ corresponds to the clean limit. The case of $n=1$ also corresponds to the clean limit but the energy scales as $rA$ instead of $A$ in the original clean limit $n=0$. In this paper, the effect of singular-coupling disorder is widely studied from $n=0$ to $1$ for $r=2$, $4$, and $10$. Interestingly, the appearance of this type of disorder with $r=4$ was predicted due to the presence of hydrogen cation vacancies in H$_3$LiIr$_2$O$_6$~\cite{yadav2018strong}.

\subsection{Method}

We employ the Lanczos exact diagonalization method of the model described by Eq.~\eqref{model} in the presence of disorder. A 24-site cluster with periodic boundary conditions is used [Fig~\ref{fig:model}(a)]. For each parameter set, we perform the simulations for 30-50 random realizations and take an average over them if needed. This setup is sufficient to investigate the instability of Kitaev QSL around the Kitaev limit because long-range magnetic orderings are not considered as discussed below. We calculate the ground-state energy $E_0$, spin-spin correlation functions $\langle \mathbf{S}_i \cdot \mathbf{S}_j \rangle$, static spin structure factor $S({\bf Q})$, and the expectation value of flux operator $\langle W \rangle$. 

Peak positions in the second derivative of ground-state energy with respect to particular parameters provide valuable information on the position of phase boundaries. Also, the width of the peaks may reflect the 'sharpness' of phase transitions; roughly speaking, a sharp (broad) peak indicates a first-order (second-order or continuous) transition. To identify the magnetic structure for each phase, we calculate the static spin structure factor
\begin{equation}
S({\bf Q})=\sum_{ij} \langle {\tilde {\bf S}}_i\cdot {\tilde {\bf S}}_j \rangle \exp[i {\bf Q}\cdot ({\bf r}_i-{\bf r}_j)],
\end{equation}
where $N$ is the number of sites in the periodic cluster ($N=24$) and ${\bf r}_i$ is the position of site $i$. For ordered phases the magnetic structure can be determined from the reciprocal-space Bragg-peak positions. The Brillouin zone and magnetic Bragg peaks for zigzag and stripy states are shown in Fig ~\ref{fig:model}(b). Note that in our calculations for the clean limit any spatial and spin symmetries are not broken, so that the structure factor exhibits all Bragg peaks for three ordering directions. However, as shown below, the translation and rotation symmetries can be broken when the disorder is introduced. In QSL phase $S({\bf Q})$ is nearly structureless and the intensity is relatively small for ${\bf Q}$'s. Furthermore, to see how close is the system to the Kitaev limit, we calculate the expectation value of flux operator. The flux operator is defined on a hexagonal plaquette by 
\begin{equation}
    W_p = 2^6 S_1^xS_2^yS_3^zS_4^xS_5^yS_6^z,
    \label{fluxop}
\end{equation}
where a possible numbering of sites is denoted in Fig~\ref{fig:model}(a). In the Kitaev limit, the flux operator is a conserved quantity, commuting with the Hamiltonian, having values exactly $\pm1$. Away from the Kitaev QSL regime, this value drops to zero. Therefore, we can also use this quantity as a measure of overlap between an observed QSL and ideal Kitaev QSL.

\section{Results}

 \begin{figure}[tbh]
	\hspace{-0.5cm}
	\includegraphics[width=\columnwidth]{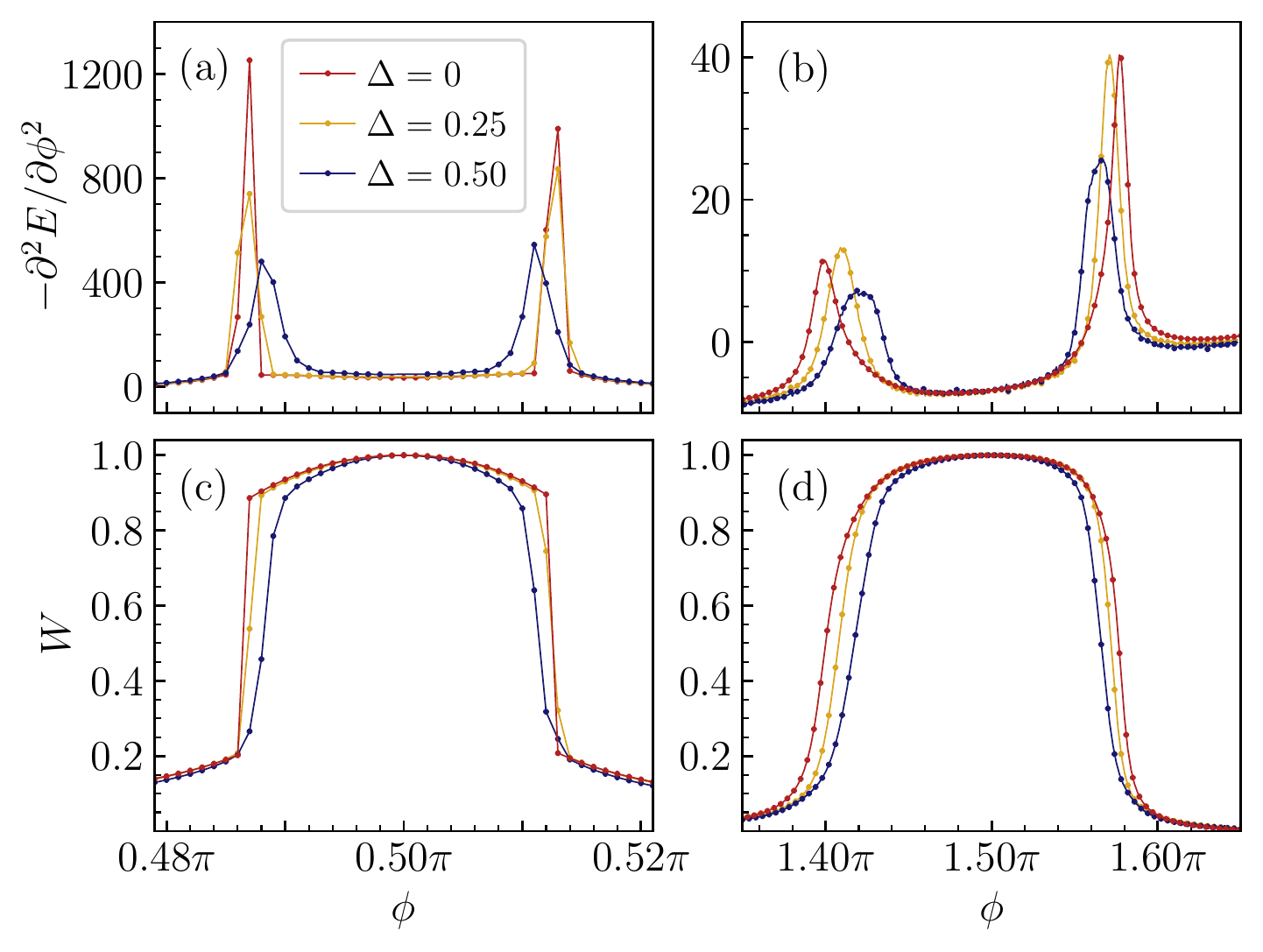}
	\caption{Second derivative of the ground-state energy as a function of $\phi$ around (a) AFM ($\phi=0.5\pi$) and (b) FM ($\phi=1.5\pi$) Kitaev points. Averaged expectation value of the plaquette operator over all 12 plaquettes around (c) AFM and (d) FM Kitaev points. For the disordered cases both the quantities are averaged over 50 disorder realizations.}
	\label{fig:dis_A}
\end{figure}

\subsection{Random-coupling disorder}\label{sec:rcd}

We begin by discussing the consequence of the random-coupling disorder. Since we are particularly interested in the possible emergence of QSL by disorder, we focus on the parameter regions around the Kitaev points $\phi=0.5\pi, 1.5\pi$; where any magnetic order can be avoided due to the fully dominant Kitaev interactions over the Heisenberg ones. Fig.~\ref{fig:dis_A}(a,b) show second derivative of the ground-state energy with respect to $\phi$, i.e., $-\partial^2E_0/\partial\phi^2$, as a function of $\phi$ for several values of the disorder strength $\Delta$ around the AFM ($\phi=0.5\pi$) and FM ($\phi=1.5\pi$) Kitaev points. 
Similarly, expectation values of the flux operator \eqref{fluxop}, i.e., $\langle W_p \rangle$, are plotted in Fig.~\ref{fig:dis_A}(c,d). In the presence of disorder, the averaged values of $-\partial^2E_0/\partial\phi^2$ and $\langle W_p \rangle$ are taken over 50 random realisations.

Around AFM Kitaev point, assuming that a peak position in $-\partial^2E_0/\partial\phi^2$ vs. $\phi$ indicates a critical point, a transition from N\'eel to Kitaev QSL phases occurs at $\phi=0.486\pi (\equiv\phi_{\rm Neel-QSL})$ and a transition from Kitaev QSL to zigzag phases at $\phi=0.514 \pi (\equiv\phi_{\rm QSL-zigzag})$ in the clean limit ($\Delta=0$). We find that the phase transitions are hardly affected by the disorder strength, $\Delta=0.25$. However, with further increasing $\Delta$ to $0.5$, the deviation from the clean limit becomes prominent. The critical points are changed to $\phi_{\rm Neel-QSL}=0.488 \pi$ and $\phi_{\rm QSL-zigzag}=0.511 \pi$. Thus, the range of Kitaev QSL has shrunk by the disorder. Even more interestingly, both peaks in $-\partial^2E_0/\partial\phi^2$ at $\Delta=0.5$ are lower and broader compared to the sharp peaks indicating first-order transitions in the clean limit ($\Delta=0$). This means that the disorder makes the phase transitions more crossover-like. It is also confirmed by a rather blunt behavior of $\langle W_p \rangle$ around the critical points.
Accordingly, the system may be characterized by a short-range N\'eel phase in a range with peak width around $\phi=\phi_{\rm Neel-QSL}$ and by a mixture of small zigzag domains and QSL regions in a range with peak width around $\phi=\phi_{\rm QSL-zigzag}$. The possibility of the formation of zigzag domains is discussed below. On the other hand, N\'eel domain walls would be not created because the random-coupling disorder does not lift the degeneracy of spin-rotation symmetry unlike that of spatial-rotation symmetry.

The effect of disorder around FM Kitaev point seems to be even more sensitive than around AFM Kitaev point. A transition from FM to Kitaev QSL phases occurs at $\phi=1.40 \pi (\equiv\phi_{\rm FM-QSL})$ and a transition from Kitaev QSL to stripy phases at $\phi=1.58 \pi (\equiv\phi_{\rm QSL-stripy})$ in the clean limit; whereas, they are $\phi_{\rm FM-QSL}=1.42 \pi$ and $\phi_{\rm QSL-stripy}=1.56 \pi$ at $\Delta=0.5$. Although the shifts of critical $\phi$ values may look tiny, we can find them somewhat considerable by thinking in terms of $K/J$ as follows: The condition to achieve a QSL is modified from $K/J>3.1$ or $K/J<-3.9$ in the clean limit to $K/J>3.9$ or $K/J<-5.2$ at $\Delta=0.5$. Nevertheless, 
we can say that the effect of the random-coupling disorder on magnetic order is relatively weak, which is consistent with the previous study~\cite{Laflorencie2006Random}. Then, as in the case around AFM Kitaev limit, peaks in $-\partial^2E_0/\partial\phi^2$ are broadened. Thus, the system may be characterized by a short-range FM phase in a range with peak width around $\phi=\phi_{\rm FM-QSL}$ and by a mixture of small stripy domains and QSL regions in a range with peak width around $\phi=\phi_{\rm QSL-stripy}$. The possibility of the formation of stripy domains is discussed below.

It is also interesting that $|\langle W \rangle|$ keeps nearly $1$ in the both AFM and FM Kitaev QSL phases despite the presence of sizeable disorder. In the Kitaev limits $\phi=\pm0.5\pi$ ($K=\pm1$, $J=0$), the flux operator is still commutative with the Hamiltonian even for finite $\Delta$. Thus, as confirmed in Fig. \ref{fig:dis_A}(c,d), the ground state is still characterized by $|\langle W \rangle|=1$ ($\forall$ hexagonal plaquette). Up to $\Delta=0.5$, we found $\langle W \rangle$ to be the unity for all studied disorder samples. Furthermore, as naively expected, we find that the nearest-neighbour spin-spin correlations are finite and longer-range ones are zero as is the case with the clean limit. Nevertheless, the value of the nearest-neighbour correlations can deviate from the clean limit value $|\langle \mathbf{S}_i \cdot \mathbf{S}_j \rangle_{\rm NN}|=0.1323$ depending on the coupling value, namely,  $|\langle \mathbf{S}_i \cdot \mathbf{S}_j \rangle_{\rm NN}|>0.1323$ for the bond with $\delta A>0$ and  $|\langle \mathbf{S}_i \cdot \mathbf{S}_j \rangle_{\rm NN}|<0.1323$ for the bond with $\delta A<0$.

\begin{figure}[tbh]
	\includegraphics[width=\columnwidth]{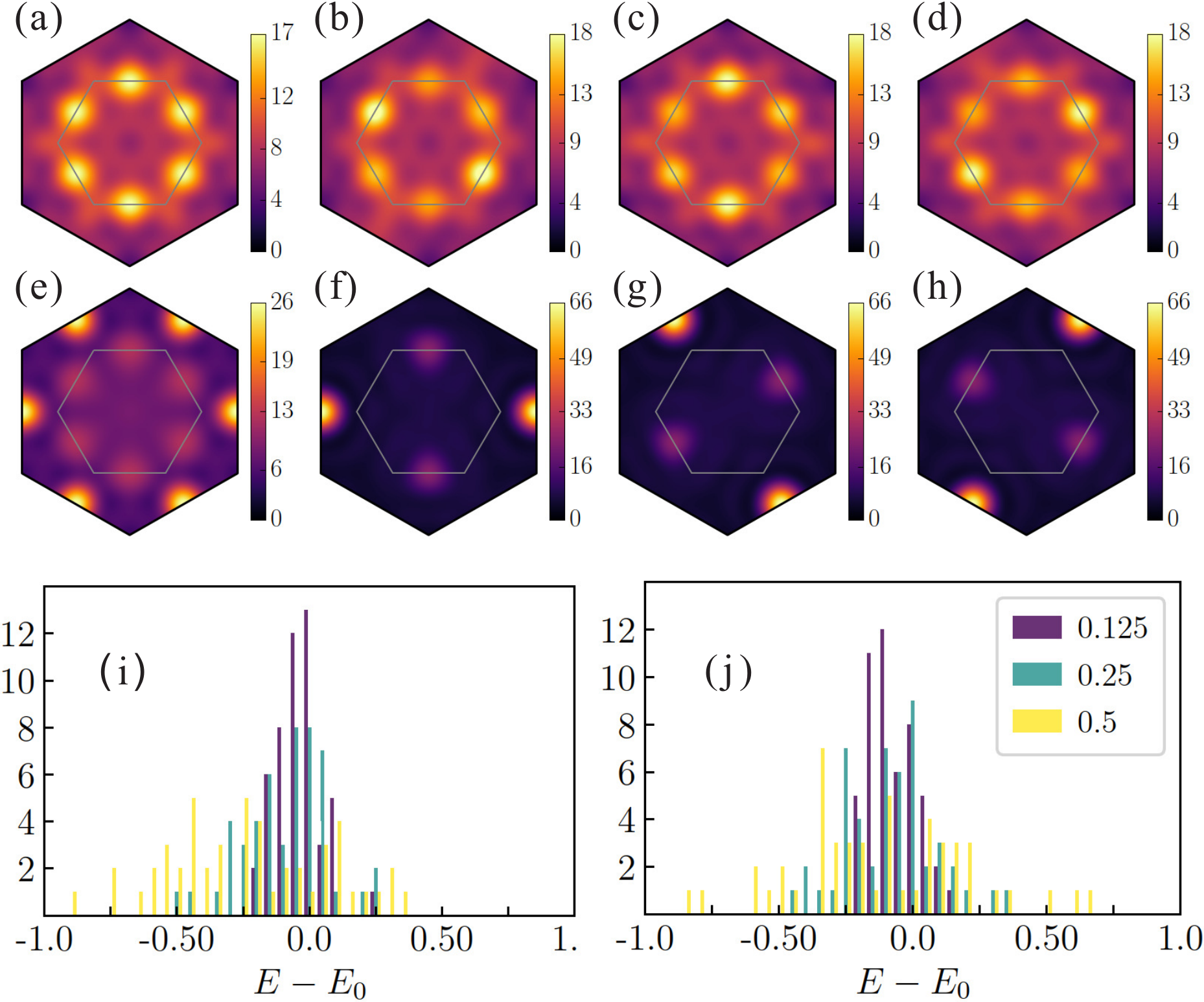}
	\caption{Static structure factors for (a) zigzag ($\phi=0.52\pi$) and (e) stripy ($\phi=1.65\pi$) states in the clean limit ($\Delta=0$). Examples of static spin structure factors with rotation-symmetry breaking for (b-d) zigzag and (f-h) stripy states at $\Delta=0.5$. (i,j) Energy distributions of 50 disorder samples for $\phi=0.52\pi$ and $\phi=1.65\pi$ at $\Delta=0.5$.}
	\label{fig:dis_A_2}
\end{figure}

Let us next consider how the ordered phases are affected by the random-coupling disorder. In the clean limit, the zigzag and stripy ordered states are associated with spatial rotation symmetry breaking. Yet, the rotation symmetry is not broken in our calculations with 24-site periodic cluster and three states with different ordering directions are energetically degenerate. Fig.~\ref{fig:dis_A_2}(a,e),  show that the Bragg peaks for all three ordering directions in zigzag and stripy order appear with the same intensity in the static structure factor. One of the degenerate three states is then selected as a ground state when its global magnetic ordering is constructed. However, the rotation symmetry can be explicitly broken once the random-coupling disorder is introduced. Examples of static spin structure factors with broken rotation symmetry for the zigzag state ($\phi=0.52\pi$) and stripy state ($\phi=1.65\pi$) are plotted in Fig.~\ref{fig:dis_A_2}(b-d) and (f-h), respectively. In a disordered system, one of the symmetry-broken state with lowest energy tends to be locally realized depending on the disorder distribution. This means in the ordered phase, the three domains with different ordering directions can coexist as a result of disorder. In Fig.~\ref{fig:dis_A_2}(i,j) their energy distributions of 50 disorder samples are shown. For both of $\phi=0.52\pi$ and $1.65\pi$, the variance of energy distribution is increased with increasing $\Delta$. The local ordering structure with lower energy is strongly stabilized and that with higher energy is easily excluded. Thus, the averaged area of zigzag (or stripy) domains becomes smaller and the system is in more like a spin-glass state for larger $\Delta$.
It is because the short-range correlation is dominated only by the local disorder structure when the disorder is strong.
In addition, a coexistence of four kinds of domains could be possible in the crossover regions between QSL and magnetic ordered phases, namely, one QSL domain and three ordering domains as mentioned above \cite{andrade2014magnetism}.

\subsection{Singular-coupling disorder}\label{sec:scd}

 \begin{figure}[tbh]
	\includegraphics[width=\columnwidth]{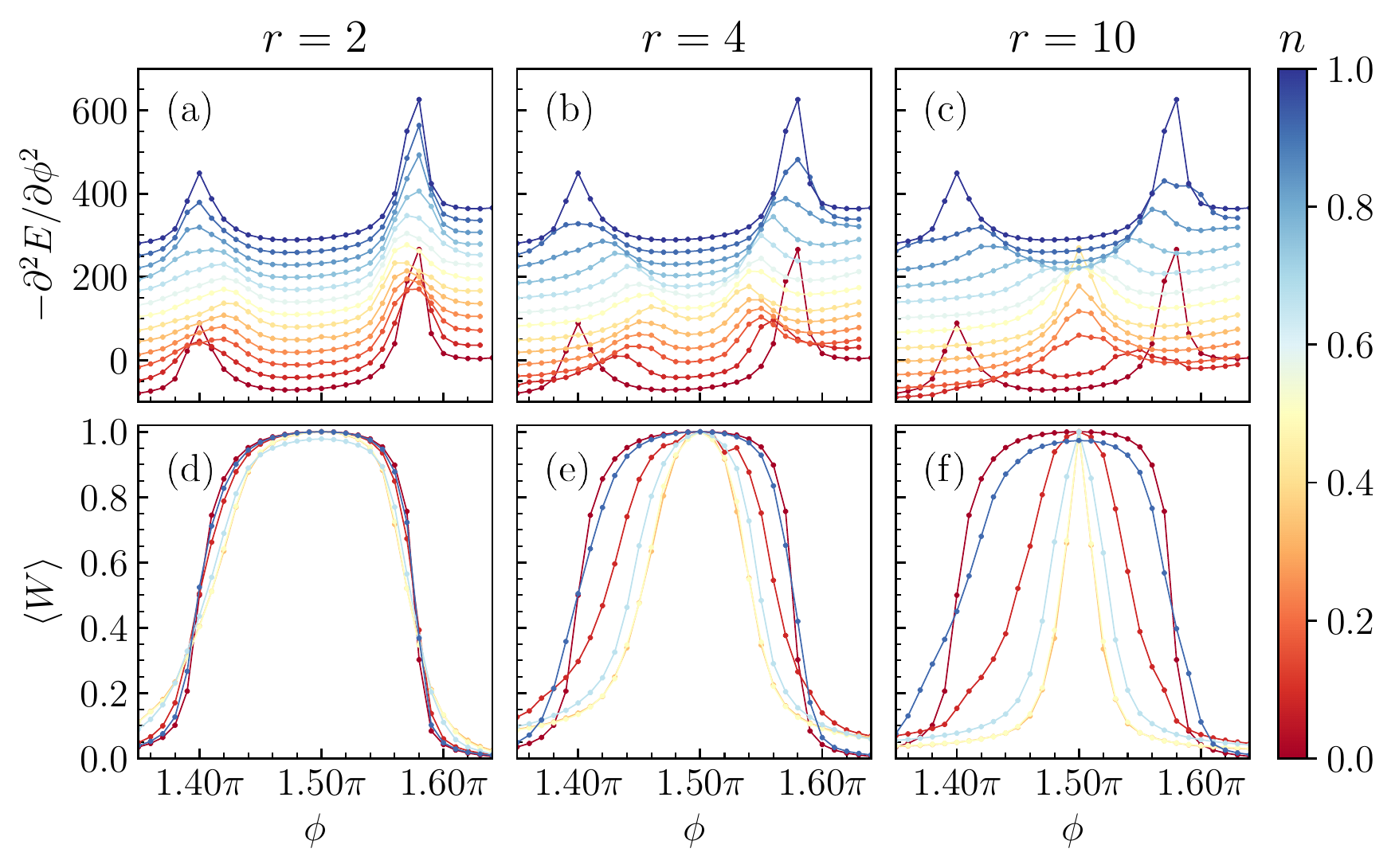}
    \caption{(a-c) Second derivative of the ground-state energy and (d-f) expectation value of the flux operator \ref{fluxop} as a function of $\phi$ around FM Kitaev point ($\phi=1.5\pi$). Results are shown for various disorder strength $r=2$, $4$, and $10$, and disorder density $n=0-1$.}
    \label{fig:dis_r}
\end{figure}

Next, we turn to the case of singular-coupling disorder. For this case we can study the effect of disorder as functions of the disorder strength $r$ and the disorder density $n$. As in the case of random-coupling disorder, the impact of singular-coupling disorder on the two phase transitions around the FM Kitaev point $\phi=1.5\pi$ is investigated by estimating $-\partial^2E_0/\partial\phi^2$ with $\phi$. Here, it is convenient to renormalize the energy unit as $A \rightarrow A/(nr+(1-n))$. In Fig.~\ref{fig:dis_r}(a-c), we plot $-\partial^2E_0/\partial\phi^2$ as a function of $\phi$ with fixed $r=2$, $4$, and $10$. The disorder density $n$ is varied from $0$ to $1$ at intervals of $3/36$. Results shown in Fig.~\ref{fig:dis_r} are averaged values of $-\partial^2E_0/\partial\phi^2$ and $\langle W_{\rm p} \rangle$ over 30 disorder samples. Interestingly, $-\partial^2E_0/\partial\phi^2$ is nearly symmetric about $n=18/36$ for any $r$. This means that the qualitative properties are determined only by the ratio of distinct bonds. At $n=0$ and $n=1$, where the system is clean, the range of FM QSL is indicated by two sharp peaks at $\phi=1.40\pi$ and $1.58\pi$. The transitions to FM ($\phi \lesssim 1.40\pi$) and stripy ($\phi \gtrsim 1.58\pi$) phases are of the first order.

For weak disorder strength ($r=2$), we find that the QSL range is slightly reduced in the whole density region $0<n<1$ and it is narrowest at $n=0.5$. This reduction seems to increase continuously with increasing $r$. As a result, for strong disorder strength ($r=10$), the QSL range almost vanishes and a nearly direct transition between FM and stripy phases may occur around $n=0.5$. The remaining QSL seems to be still of the Kitaev-type because of $\langle W_p \rangle \approx 1$. Another effect of singular-coupling disorder is that the peak of $-\partial^2E_0/\partial\phi^2$ indicating a transition is significantly broadened even for weak disorder. This implies that peak position for single disorder sample differs from each other. Thus, transition in the disordered system becomes continuous or even a crossover-like one. As in the case of random-coupling disorder, the system may be characterised by a short-range FM or a short-range stripy ordering in the crossover region. This might be interpreted as a QSL in the broad sense of the term. It is rather surprising that the broadening of peak in $-\partial^2E_0/\partial\phi^2$ seems to be almost independent of $r$. 

 \begin{figure}[tbh]
	\includegraphics[width=\columnwidth]{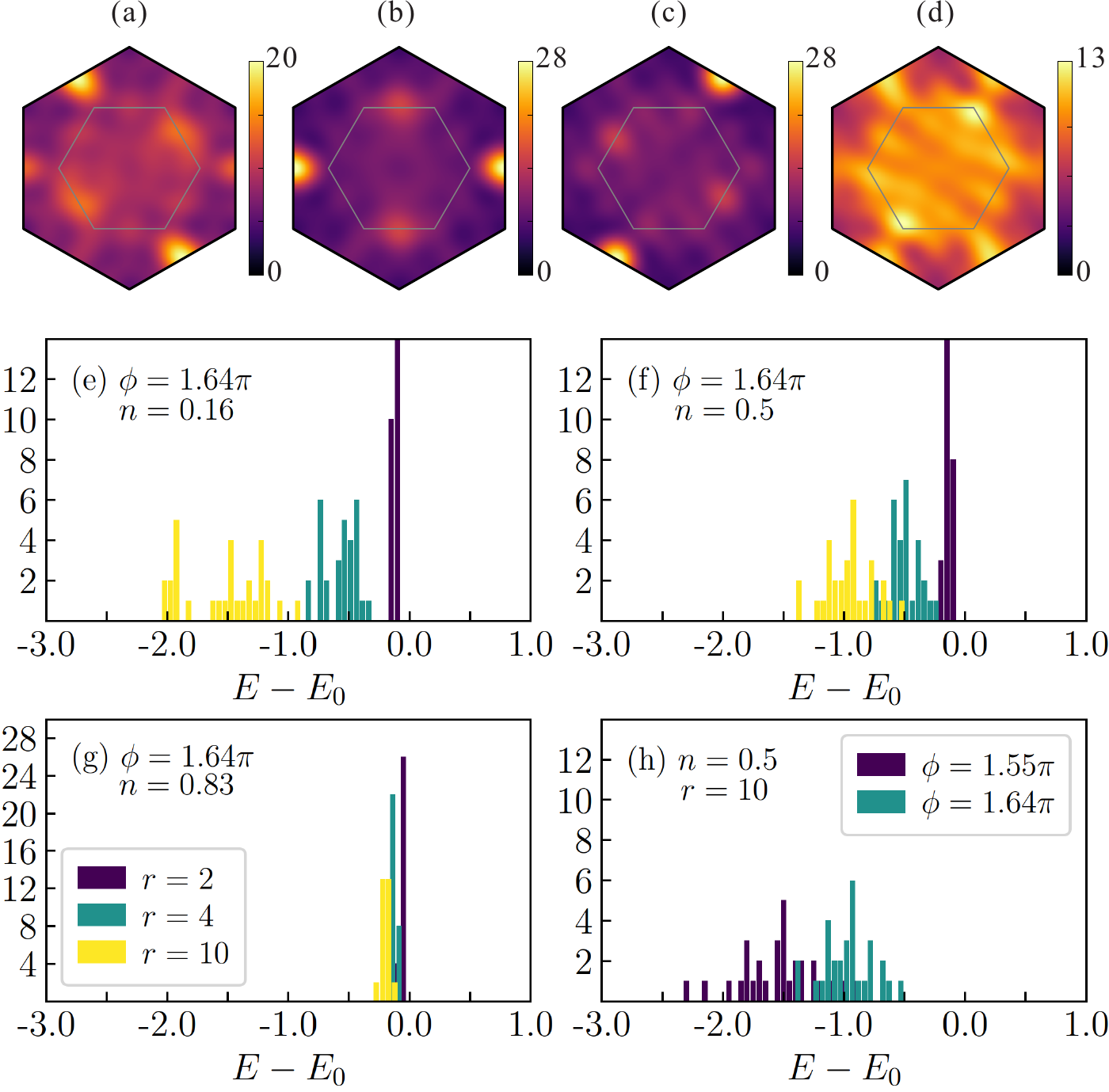}
	\caption{(a-c) Representative static structure factors of rotation-symmetry broken stripy ordering at $\phi=1.64\pi$, $n=0.5$ for $r=10$. (d) QSL-like structure factor appearing at $\phi=1.55\pi$, $n=18/36$ for $r=10$. (e-h) Energy distributions of 30 samples for various parameters, where the energy value for the clean limit is set to be zero and the class interval is $\Delta E_0=0.05$.
	}
	\label{fig:energy_dist}
\end{figure}

Furthermore, in order to see how the ordered state is affected by the singular-coupling disorder, we study the static structure factor in the stripy phase. Three representative structure factors at $\phi=1.64\pi$, $n=0.5$ for $r=10$ are shown in Fig.~\ref{fig:energy_dist}(a-c), indicating a stripy state with broken rotation symmetry. Since most of samples for any $r$ basically exhibit one of the three structures, the system consists of the three stripy domains with different ordering directions. As discussed in Sec.~\ref{sec:rcd}, the variation of energy distribution is related to the averaged area of stripy domains. In Fig.~\ref{fig:energy_dist}(e-g) the energy distributions of 30 disorder samples in the stripy phase ($\phi=1.64\pi$) are shown. As expected, the variance of energy distribution increases with increasing $r$, and however, it clearly decreases with increasing $n$. As a result, their energy distributions are completely different between small and large $n$ regions, although the second derivative of energy is symmetric about $n=0.5$. This may be explained by considering random-singlet formations. At lower $n$, the energy can be easily lowered because the stronger $r$ bonds are mostly isolated and spin-singlets are formed on them. The  direction of stripy ordering is closely related to the distribution of random-singlets. Therefore, the area of stripy domains tends to be small at smaller $n$ [Fig.~\ref{fig:energy_dist}(e)], and it would be relatively large at larger $n$ [Fig.~\ref{fig:energy_dist}(g)]. For example, at $n=30/36$ the variance seems to be very small and the energies of disorder samples are almost degenerate for any $n$. This situation is close to the clean limit. Thus, the system consists of large stripy domains if they exist. On the other hand, at $n=6/36$ and $r=10$ a wide distribution of the variance may indicate the appearance of small stripy domains. This is because a local stripy configuration can be realised against a global ordering if it significantly lowers the total energy. In sum, a spin-glass state due to a mixture of tiny stripy domains may be achieved at large $r$ and small $n$.

Let us look at the crossover region between stripy and Kitaev QSL phases in a little more detail. We take the case of $r=10$ and $n=18/36$ for example. A structureless structure factor as in Fig.~\ref{fig:energy_dist}(d) appears other than the three symmetry broken ordered ones. Also, the variation of energy distribution is relatively larger than that in the deep stripy phase [Fig.~\ref{fig:energy_dist}(h)]. Thus, a coexistence of four kinds of domains may be possible in the crossover region between Kitaev QSL and stripy phases, namely, one QSL domain and three stripy ordering domains. At $\phi=1.55\pi$, 10 of 30 samples exhibit such a QSL-like structure factor, and the remaining 20 samples exhibit either of stripy ones. Assuming that $\langle W \rangle = 0$ for the stripy samples and $\langle W \rangle = 1$ for the QSL-like samples, we may roughly expect $\langle W \rangle \sim 0.33$. This is comparable to the actual value $\langle W \rangle \sim 0.4$ for $\phi=1.55\pi$ at $r=10$ and $n=18/36$.

\begin{figure}[tbh]
	\includegraphics[width=1.0\columnwidth]{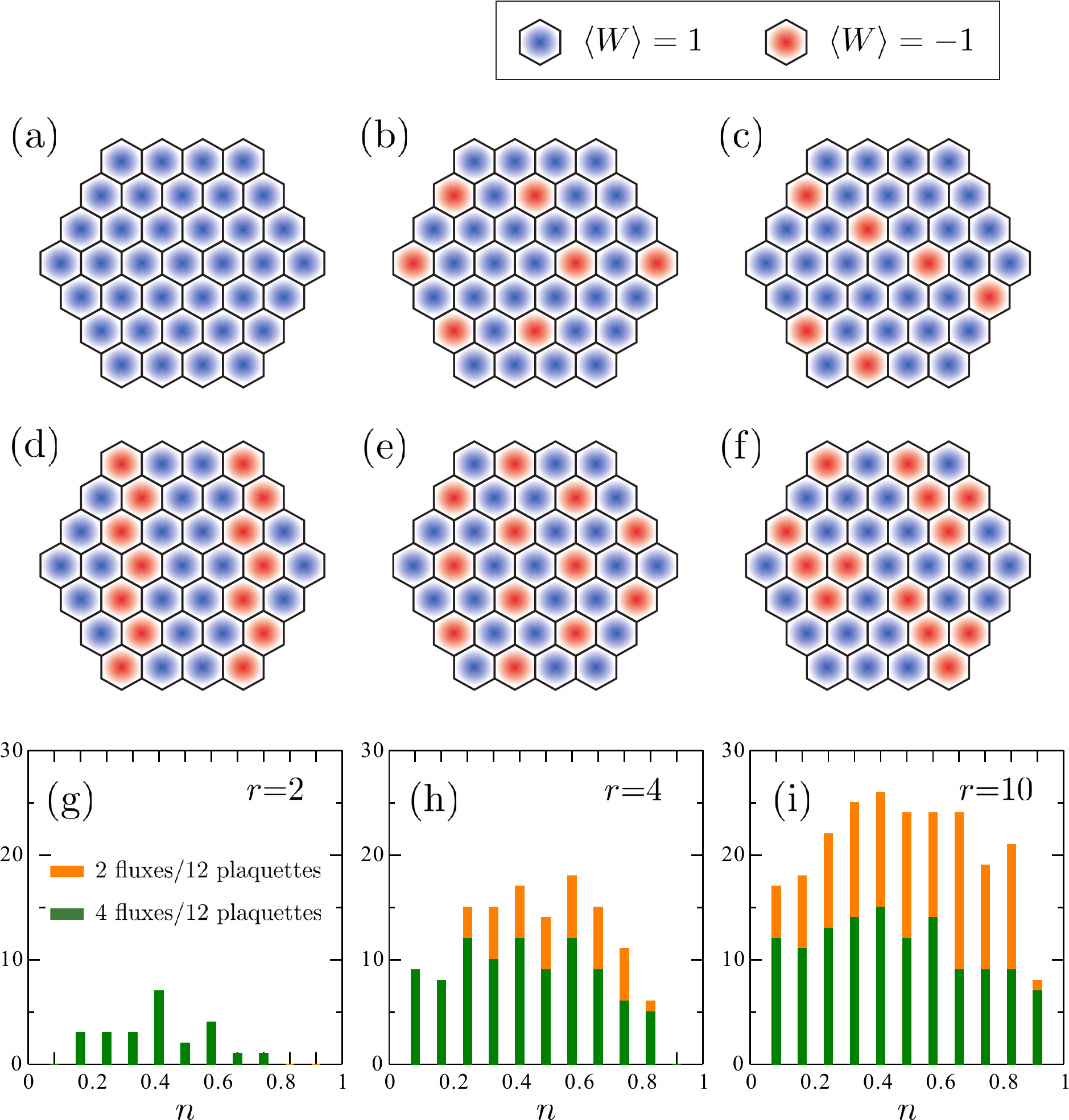}
	\caption{Typical examples of flux configurations for (a) flux-free, (b,c) 1/6 flux density, and (d-f) 1/3 flux density states. Number of samples exhibiting flux states out of 30 samples, as a function of disorder density $n$ for (g) $r=2$, (h) $r=4$, and (i) $r=10$}.
	\label{fig:flux}
\end{figure}

\subsubsection*{Appearance of flux state by disorder}

We find an intriguing evolution of flux states with the disorder strength $r$ and disorder density $n$ in the Kitaev limit $\phi=3\pi/2$. The introduction of disorder may mimic thermal ﬂuctuations in real materials. Since the relations $W_p^2=1$, $[\mathcal{H},W_p]=0$, and $[W_p,W_{p'}]=\delta_{pp'}$ are fulfilled at $\phi=3\pi/2$ even in the presence of disorder, Therefore, the eigenstate for the system~\eqref{model} can be still characterized by a set of $\langle W_p \rangle$ for all plaquettes, where $\langle W_p \rangle$ has a value of 1 or -1. For each disorder sample we see the $\langle W_p \rangle$ values for 12 plaquettes in our 24-site periodic cluster.

For the clean case ($r=1$), it is known that the ground state is given by $\langle W_p \rangle = 1$ ($\forall p$), the so-called flux-free state [Fig.~\ref{fig:flux}(a)]. Then, the excited flux states can be created by flipping $\langle W_p \rangle$ from $1$ to $-1$ for even number of plaquettes. The flux gap has been estimated to be $E_{\rm gap}=0.066K$ as a lowest-energy excitation when the values of $\langle W_p \rangle$ for two neighboring plaquettes are flipped~\cite{zschocke2015physical}.  The ﬂux density is deﬁned by $n_{\rm F}=N_{\rm F}/N_{\rm p}$, where $N_{\rm F}$ is the numbers of plaquettes exhibiting $\langle W_p \rangle=-1$ and $N_{\rm p}$ is the total number of plaquettes. In particular, the flux density $n_{\rm F}=1/6$ is achieved by a 24-site periodic cluster containing 2 plaquettes with $\langle W_p \rangle=-1$ and 10 plaquettes with $\langle W_p \rangle=1$, i.e., $N_{\rm F}=2$ and $N_{\rm p}=12$. Similarly, $n_{\rm F}=1/3$ corresponds to $N_{\rm F}=4$ and $N_{\rm p}=12$ in our calculations. We refer to $n_{\rm F}=1/6$ and $n_{\rm F}=1/3$ as 2-flux and 4-flux state, respectively. Typical examples of flux configurations for (excited) 2-flux and 4-flux states 
are illustrated in Fig.~\ref{fig:flux}(b-f).
The appearance of fluxes in the ground state implies the closing of flux gap by disorder. It is consistent with the previous study \cite{dantas2022disorder}.

Flux states appears as a ground state on introduction of singular-coupling disorder. The number of samples exhibiting flux states out of 30 samples is shown as a function of disorder density $n$ for several disorder strengths in Fig.~\ref{fig:flux}(g-i). For a weak disorder strength ($r=2$) only $\sim 10-20\%$ of disorder samples exhibit flux states. Interestingly, all of them are the 
4-ﬂux states shown in Fig.~\ref{fig:flux}(e). Since the effect of disorder is most highlighted around $n=0.5$, we  find the probability of observing a flux state is also maximum around $n=0.5$. With further increasing $r$ up to $4$ the probability of finding flux states increases. The 4-flux configuration still accounts for a large fraction of observed flux states, although a different pattern for 4-flux configurations such as in Fig.~\ref{fig:flux}(f) as well as the 2-flux configuration like in Fig.~\ref{fig:flux}(b,c) appear as minority. In other words, ordered 4-flux state would tend to be energetically 
more favourable than the other flux states. This may be related to the “Majorana insulating” state~\cite{Koga2021Majorana}. Looking at all of the flux configurations, the fluxes seem to keep distance from each other. Thus, a kind of repulsion may exist between fluxes, and the energetically-favoured ordered-flux state could be a consequence of the repulsive interaction (also see the Appendix \ref{sec:app_a}, \ref{sec:app_b}).
This also implies no possibility of a phase-separation-like flux configuration induced by disorder.
Note that, however, the ordered-flux state is not globally stabilized 
in our disordered system because various flux patterns locally exist depending on the disorder realization, for example, as shown in in Fig.~\ref{fig:flux}(b-f).

For a very strong disorder strength ($r=10$) the 4-flux and 2-flux states appear with similar probability, and the flux-free state still exists. Furthermore, various flux configurations are mixed. For example, a periodic configurations like in Fig.~\ref{fig:flux}(f) are also contained. These results indicate a trend to the random-flux states~\cite{zschocke2015physical,Kao2022spin_gap}. This can be naively understood if we can assume that the effect of disorder is essentially the same as thermal ﬂuctuations. 
Our focus here has been only the ground state, we show that random flux state appears as a consequence of disorder itself, which indicates that the flux gap closes at some disorder strength, $r$ for a particular disorder configuration.  Eventually, the presence of disorder in a random flux sector can explain the abundance of low energy states in H$_3$LiIr$_2$O$_6$ compound~\cite{kimchi2018scaling}. Thus, the low-lying flux and spin excitations~\cite{Song2016Low} of the Kitaev-Heisenberg model should be studied in future.

\subsection{Further neighbor interactions}

 \begin{figure}[tbh]
    \includegraphics[width=\columnwidth]{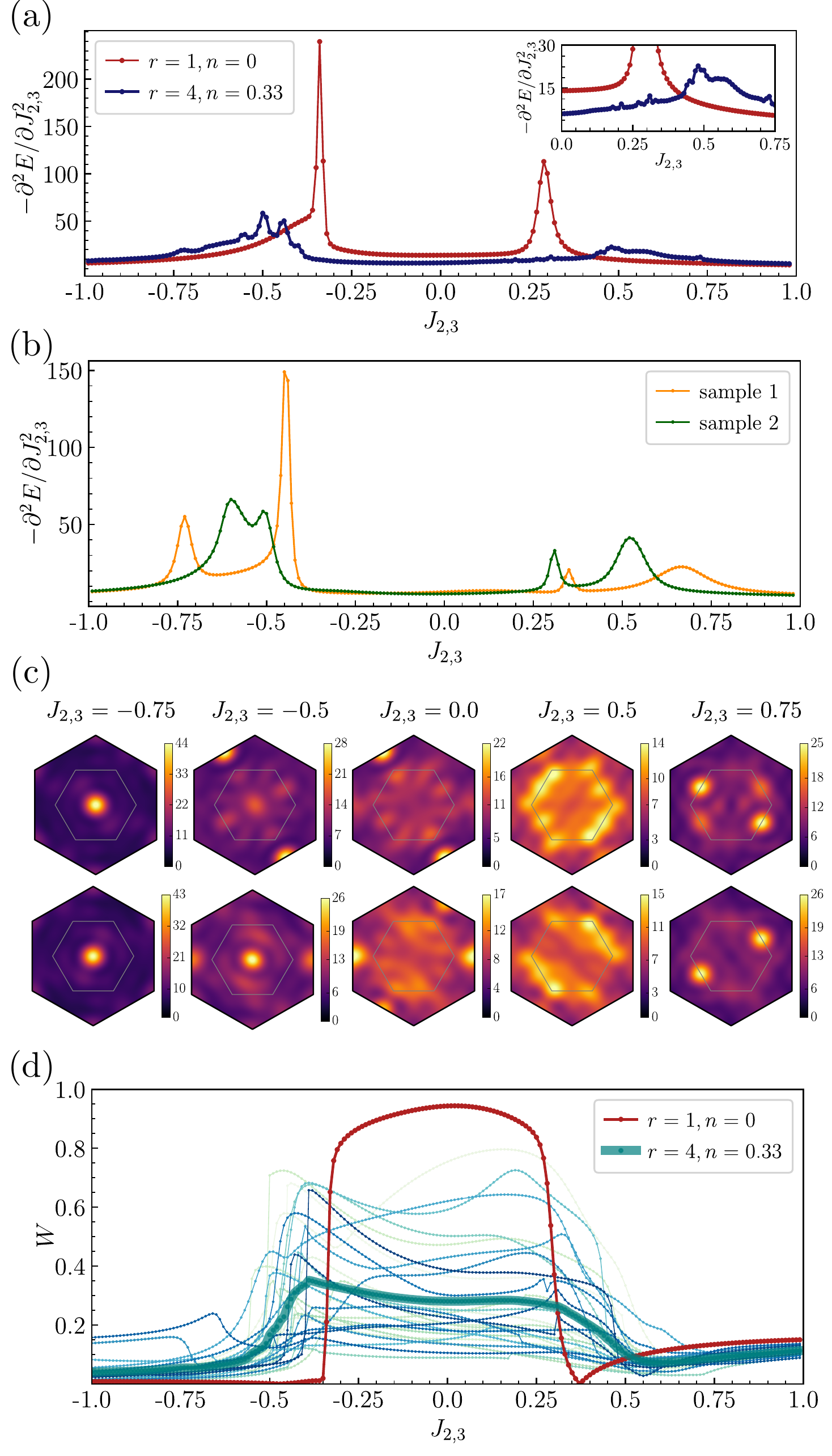}
    \caption{(a) Second derivative of the ground-state energy as a function of $J_{2,3}$ for the clean and disordered systems. The values for the system is averaged ones over 30 disorder samples. (b) Second derivative of the ground-state energy as a function of $J_{2,3}$ and (c) the corresponding static spin structure factors for two representative disorder samples. (d) Expectation values of the flux operator, $\langle W_{\rm p} \rangle$, as a function of $J_{2,3}$ for the clean and disordered systems. For disordered systems, the values for each disorder sample are also plotted.
}
    \label{fig:j2j3}
\end{figure}

\begin{figure*}[tbh]
    \includegraphics[width=\textwidth]{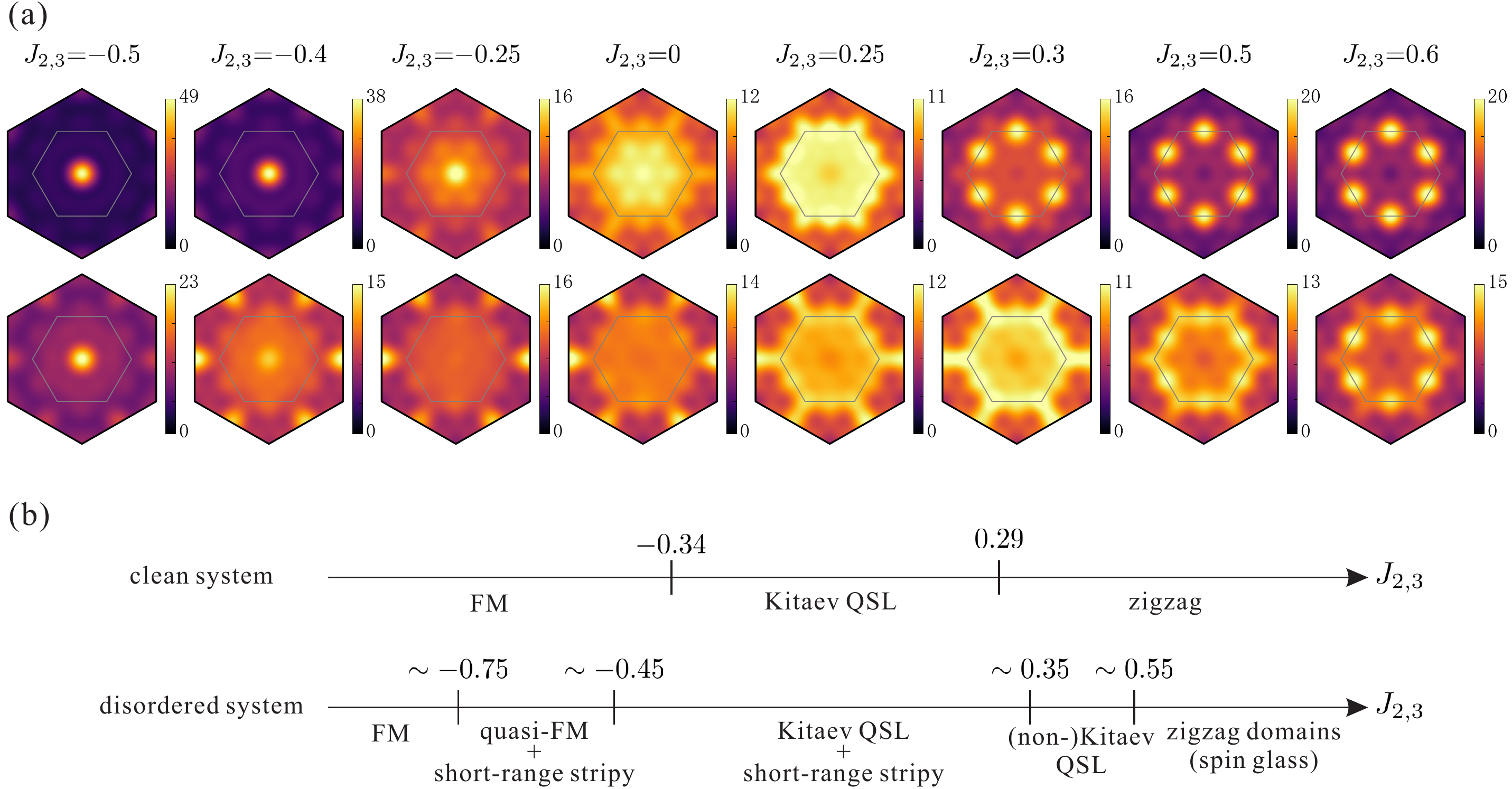}
    \caption{(a) Static spin structure factors for the clean case (upper panel) and the disordered case (lower panel) for various values of $J_{2,3}$, where $K=-6$ and $J=1$ are fixed. (b) Possible ground-state phase diagrams for the clean and disordered systems as a function of $J_{2,3}$.}
    \label{fig:j2j3-pd}
\end{figure*}

So far we have considered the effect of disorder on quantum phase transitions between Kitaev QSL and magnetic ordered states in the Kitaev-Heisenberg model [Eq.~\eqref{model}], where the ordered states are caused by the competition of nearest-neighbor exchange coupling $J$ and Kitaev term $K$. However, in real materials like Na$_2$IrO$_3$ and $\alpha$-RuCl$_3$, the experimentally observed zigzag ordering may be possibly stabilised by long-range neighbor couplings $J_2$ and $J_3$  \cite{yadav2016kitaev}. Therefore, it would be informative to study the effect of disorder on zigzag state stabilised by $J_2$ and $J_3$. Thus, the Hamiltonian is modified as
\begin{equation}
\mathcal{H}=\mathcal{H}_{\rm KH}+\sum_{(i,j)} \sum_{\alpha=2,3} J_{\alpha} \mathbf{S}_i \cdot \mathbf{S}_j
\label{modelj2j3}
\end{equation}
where $J_2$ and $J_3$ are uniform Heisenberg exchange interactions on 2nd and 3rd nearest neighbours, respectively. We fix $K=-6$, $J=1$ as a typical parameter set for the Kitaev materials. We omit the $\Gamma$ terms from our model~\cite{gammaterms} and set $J_2=J_3(\equiv J_{2,3})$ for simplicity. We now focus on the case of $n=1/3$, i.e., 12 bonds out of original 36 bonds are randomly replaced by singular couplings with disorder strength $r=4$. This value of $r$ was actually predicted as a singular-coupling disorder due to the missing of hydrogen cations in H$_3$LiIr$_2$O$_6$~\cite{yadav2018strong}.

Let us first see the second derivative of ground-state energy as a function of $J_{2,3}$, i.e., $-\partial^2E_0/\partial J_{2,3}^2$. The results for the clean and disordered cases are compared in Fig.~\ref{fig:j2j3}(a). In the clean case, two phase boundaries are signalled by the sharp peaks. We thus find three phases depending on the value of $J_{2,3}$: FM ($J_{2,3} \lesssim -0.34$), Kitaev-type QSL ($-0.34 \lesssim J_{2,3} \lesssim 0.29$), and zigzag ($J_{2,3} \gtrsim 0.29$). At $J_{2,3}=0$, although the ground state is a QSL, the system is in the vicinity of the stripy phase (the phase boundary is $K=-4$, $J=1$), For the disordered case, the averaged values over 30 disorder samples are plotted. Compared to the clean case, the heights of both peaks are much reduced and the widths are significantly broadened. Furthermore, it is interesting that the phase between FM and zigzag, seems to be extended by the disorder since the left (right) peak position is shifted from $J_{2,3} \approx -0.34$ ($J_{2,3} \approx 0.29$) to $J_{2,3} \approx -0.5$ ($J_{2,3} \approx 0.5$).

In order to take a more detailed look, let us see the result for each disorder sample. We find that most of disorder samples exhibit three or four peaks, i.e., the appearance of additional one or two peaks to the clean case, in $-\partial^2E_0/\partial J_{2,3}^2$. Typical examples of $-\partial^2E_0/\partial J_{2,3}^2$ for two disorder samples are plotted in Fig.~\ref{fig:j2j3}(b). For both samples there exist five `regions' separated by four peaks. Fig.~\ref{fig:j2j3}(c) shows the static spin structure factor at representative $J_{2,3}$ values for each region: $J_{2,3}=-0.75$, $-0.5$, $0$, $0.5$, and $0.75$. Incidentally, in the clean case, the system is in a FM phase for $J_{2,3}=-0.75$ and $-0.5$; in a QSL phase at $J_{2,3}=0$; and in a zigzag phase at $J_{2,3}=0.5$ and $0.75$. The two sets of $S({\bf Q})$ for disordered samples show similar qualitative features at the same $J_{2,3}$ value. At $J_{2,3}=-0.75$ we find a sharp peak at ${\bf Q}={\bf 0}$ indicating FM state, as in the clean case. At $J_{2,3}=-0.5$ a stripy structure appears in addition to the FM peak. Therefore, the state may be characterized as a quasi-FM order with short-range stripy fluctuations. Note that the position of stripy Bragg peaks reflects one of three possible ordering directions in each disorder sample. At $J_{2,3}=0$ it is interesting that a QSL-like structure, i.e., structureless distribution of the intensity over the first Brillouin zone, and a stripy Bragg peaks coexist. Since the system is in the proximity to stripy phase in the clean case, a short-range stripy correlations could be enhanced by a pinning effect with disorder. Nevertheless, 
as seen in Fig.~\ref{fig:j2j3}(d) the value of $\langle W_{\rm p} \rangle$ is still significantly larger than zero for most of the disorder samples, so that the system is broadly characterized as a Kitaev QSL. At $J_{2,3}=0.5$ the intensity is widely distributed and there is no sharp Bragg peaks indicating a specific ordering. Since the value of $\langle W \rangle$ is nearly zero for most of the disorder samples, the system is in a non-Kitaev QSL state. 
In this QSL phase, unlike a very rapid decay of the spin-spin correlation functions in the Kitaev QSL phase, the correlation survives at longer distance (see the Appendix \ref{sec:app_c} for more details).

At $J_{2,3}=0.75$ the structure factor for each disorder sample exhibits a symmetry-broken zigzag state with one of three possible ordering directions. Thus, the system consists of three zigzag domains. As discussed in Sec.~\ref{sec:scd} the size of domains depends on the disorder density and strength.
Also, it is worth mentioning that the QSL range is somewhat expanded by disorder. It may be because the disorder-induced stripy fluctuations push out the phase boundaries of neighboring ordered phases.

In Fig.~\ref{fig:j2j3-pd}(a) the evolution of the static structure factors with $J_{2,3}$ is compared for the clean and disordered case. For the disordered case, we show the structure factor averaged over ones over 30 disorder samples. Although the overall explanation has been already given above for two disorder samples,the effect of disorder can be more obviously seen in the averaged $S({\bf Q})$. Of particular interest is that the sharp Bragg peak features in the clean system are totally collapsed in the disordered system at $J_{2,3}=-0.4$, $0.3$, and $0.5$. This means that even short-range magnetic orderings are destroyed by disorder. Accordingly, the $J_{2,3}$-range of QSL phase is extended from $-0.34 \lesssim J_{2,3} \lesssim 0.29$ to $-0.45 \lesssim J_{2,3} \lesssim 0.55$ by the disorder as shown in prospective ground-state phase diagrams [Fig.~\ref{fig:j2j3-pd}(b)]. This is in a strong contrast to the results for the case of NN Kitaev-Heisenberg model, where the QSL phase is always shrank by any types of disorder. As mentioned above, this value of $r=4$ corresponds to the singular-coupling disorder in H$_3$LiIr$_2$O$_6$~\cite{yadav2018strong}, however, the actual values of $J_2$, $J_3$, and $n$ are unknown. Nevertheless, since the qualitative features would be unaffected by changing $n$, assuming typical values of $J_2$ and $J_3$ as several tens of percent of $J$ we can speculate that either spin glass or QSL is likely as low-temperature state of H$_3$LiIr$_2$O$_6$. Moreover, we find random-singlet formations between second- and third-neighbor spins in our system. This is qualitatively consistent with heat capacity measurement under magnetic field~\cite{kimchi2018scaling}. For more quantitative considerations the $\Gamma$ terms should be also taken into account because they are expected to be sizeable

\section{Conclusion}

Using Lanczos exact diagonalization technique we have studied the Kitaev-Heisenberg model on a honeycomb lattice in the presence of bond disorder. Especially, the effects of disorder on the transition between Kitaev spin liquid and magnetic ordered states as well as the stability of magnetic ordering have been investigated. We have considered two types of disorder:  random-coupling disorder and singular-coupling disorder. The former is achieved by a random distribution of bond strength, and the latter by a random mixture of two kinds of bond strengths. They exhibit qualitatively similar effects in the pure Kitaev-Heisenberg model without long-range interactions. The range of Kitaev spin liquid phase is reduced and the transition to magnetic ordered phases becomes more crossover-like. Accordingly, although the regions of zigzag and stripy phases are enlarged by disorder, their long-range orderings in the clean system are replaced by a mixture of the three domains with different ordering directions. With increasing the disorder strength the area of domains becomes smaller and the system goes into a spin-glass state. Particularly in the crossover regime, the coexistence of magnetic domains and Kitaev spin-liquid domains may be possible. On the other hand, qualitative trend is different if magnetic ordered phases caused by long-range interactions are doped by disorder. The stability of such magnetic ordering is diminished by singular-coupling disorder, and as a result, the range of spin-liquid is extended. This mechanism may be relevant to materials like $\alpha$-RuCl$_3$ and H$_3$LiIr$_2$O$_6$ where the zigzag ground state is stabilized by small long-range interactions.

We have also found that the flux states are induced by a certain amount of disorder strength. Interestingly, the fluxes tend to form kinds of commensurate ordering at intermediate disorder strength. This may be related to the Majorana insulating” state. This may indicate an existence of repulsive interaction between fluxes. With further increasing disorder strength the system goes into the random-flux states. At any disorder strength we have seen no signature for a phase-separation-like flux configuration.

\section{Acknowledgments}

We acknowledge Joji Nasu, Rajyavardhan Ray, and Pranay Patil for fruitful discussions. We thank Ulrike Nitzsche for technical assistance. This project is funded by IFW Excellence Programme 2020 and the German Research Foundation (DFG) via the projects A05 of the Collaborative Research Center SFB 1143 (project-id 247310070)
and through the Würzburg-Dresden Cluster of Excellence on Complexity and Topology in Quantum Matter ct.qmat (EXC 2147, Project No. 39085490).

\section{Appendix}

\subsection{Excitation of free-flux state by applying local flux field}
\label{sec:app_a}
\begin{figure}[h]
	\includegraphics[width=0.8\columnwidth, scale =0.5]{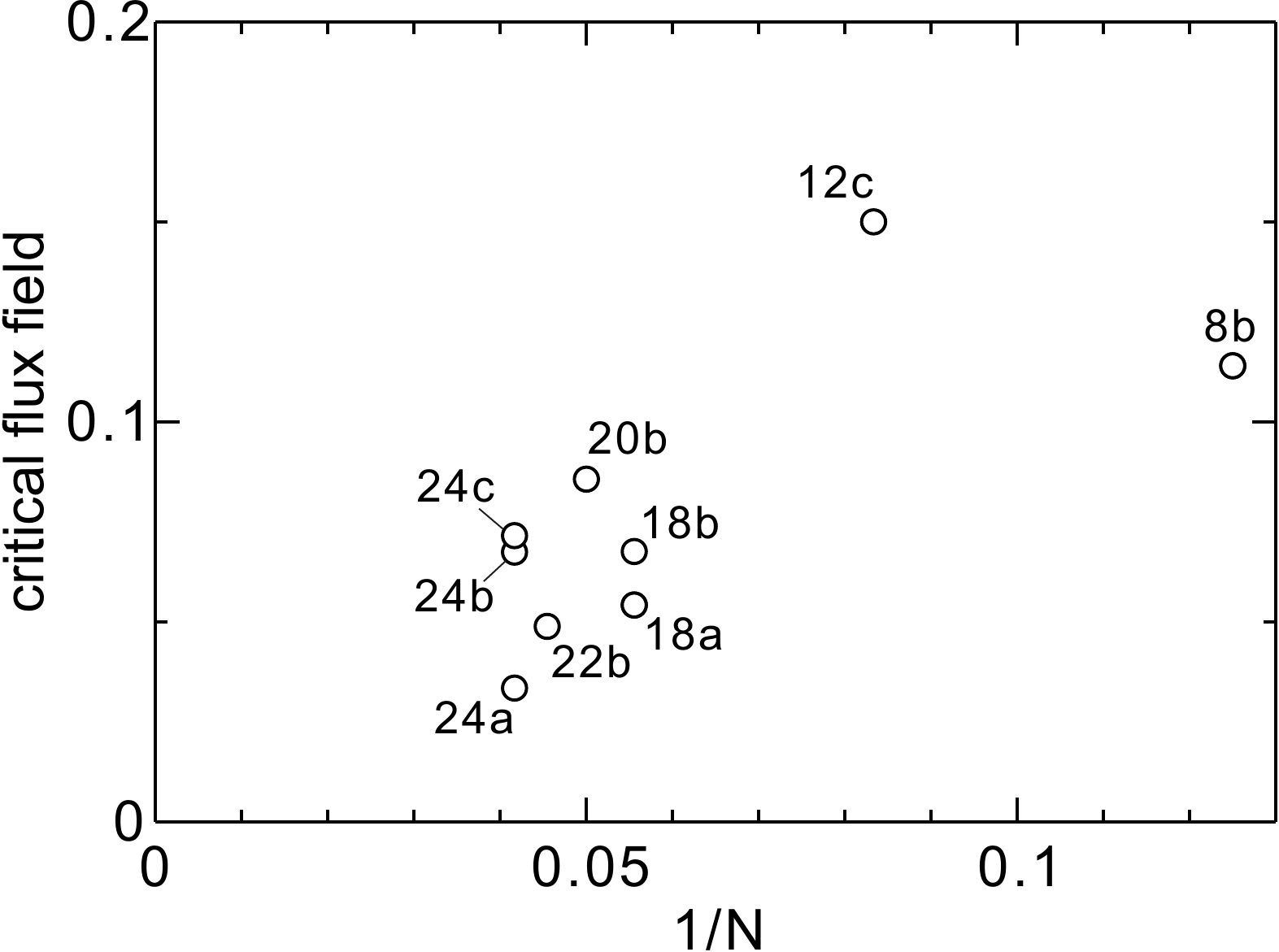}
	\caption{Critical flux field for various clusters. The used clusters are
		shown in Fig.~\ref{fig:cluster_SM}(a).
	}
	\label{fig:ff_SM}
\end{figure}

\begin{figure}[tbh]
	\includegraphics[width= 0.8\columnwidth]{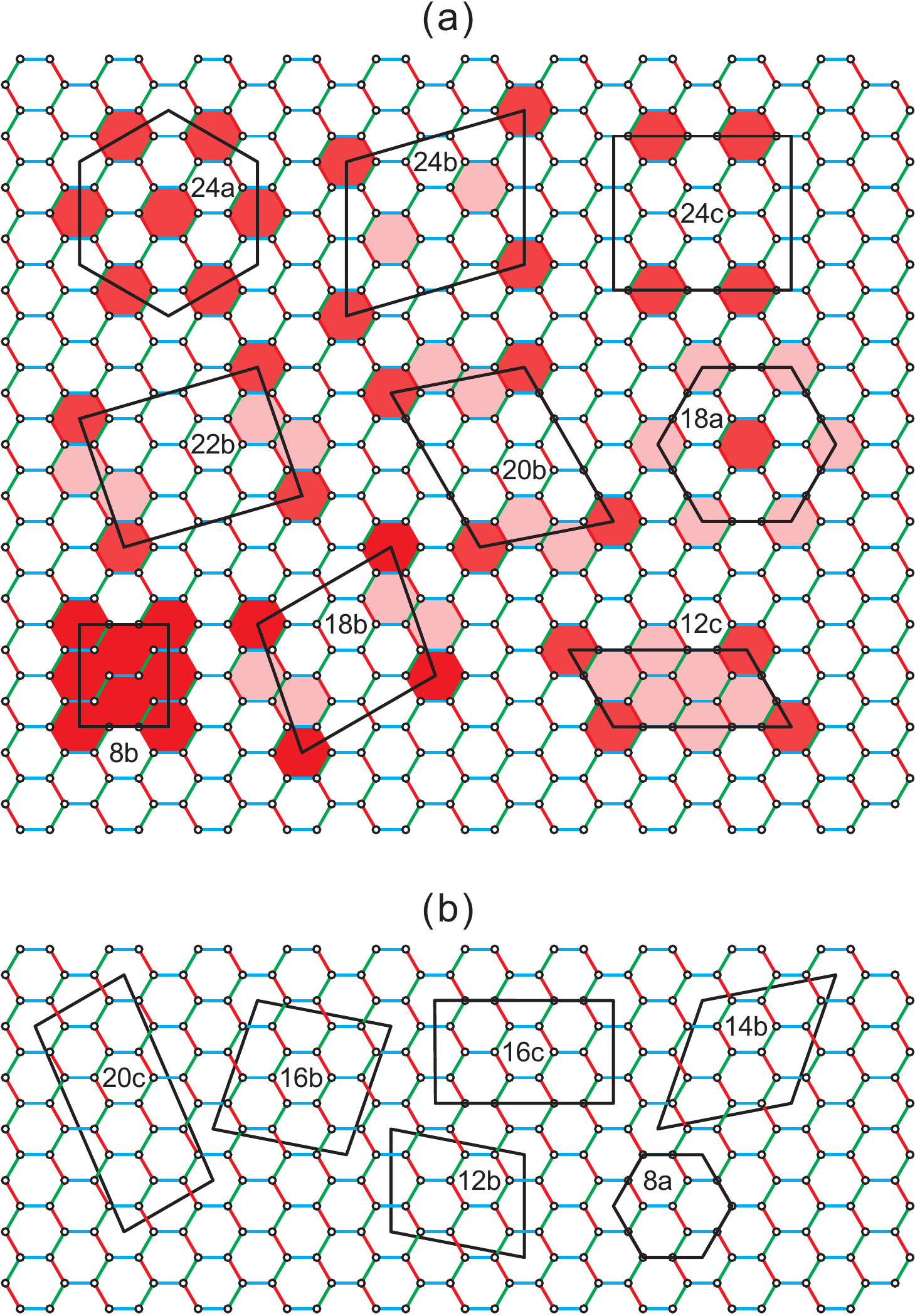}
	\caption{(a) Clusters used for estimating the critical flux field. The red
		plaquette corresponds to $\langle W_p \rangle=-1$, and either of the
		thin red plaquettes is occupied by  $\langle W_p \rangle=-1$ in the
		excited flux state for each cluster with two fluxes. (b) Clusters
		having non-flux-free ground state (see text).
	}
	\label{fig:cluster_SM}
\end{figure}

It is known that the ground state of the Kitaev model on a honeycomb lattice
is characterized by $\langle W_p \rangle = 1$ ($\forall p$), the so-called
flux-free state. The flux operator is defined as
$W_p=2^6S_1^xS_2^yS_3^zS_4^xS_5^yS_6^z$ (also see the main text).
An excited flux state can be created by flipping
$\langle W_p \rangle$ from $1$ to $-1$ for even number of plaquettes.
We here consider the Kitaev model with $K=-2$.
To study flux configuration in the excited flux state, we apply flux field
$h_{\rm f}W_p$ to a single plaquette on finite-size clusters shown in
Fig.~\ref{fig:cluster_SM}(a). This is a kind of the pinning-field method
which is usually used to study magnetically ordered state. With increasing
$h_{\rm f}$, the value of $\langle W_p \rangle$ for the field-applied
plaquette is flipped from $1$ to $-1$ at some $h_{\rm f}$. We call it critical
flux field. In Fig.~\ref{fig:ff_SM} the critical flux field is plotted as a
function of $1/N$. Above the critical flux field, the clusters 24a and 8b contain four fluxes, and the other clusters contain only two fluxes.
For the clusters with two fluxes, the position of fluxes may be not uniquely
fixed due to the symmetry of cluster. For example, let us see the cluster
24b in Fig.~\ref{fig:cluster_SM}(a). A red plaquette corresponds
to $\langle W_p \rangle=-1$ where the flux field is applied. The remaining
$\langle W_p \rangle=-1$ is put into either of two thin red plaquettes.

Interestingly, the flux configuration for 24a and 18a exhibits a staggered
order. (Note that the cluster denoted by 18a is not perfectly consistent
with a staggered order because three fluxes are needed to complete it.
Nevertheless, the possible positions of fluxes are of the staggered order
[see Fig.~\ref{fig:cluster_SM}(a)].) As seen in Fig.~\ref{fig:ff_SM},
the critical flux fields for 24a and 18a are lower than those for the
other clusters with the same system size. Especially, the critical flux
field for 24a is appreciably lower than that for24b and 24c. This may mean
that the staggered ordered state is more stable than the other flux state.
This is consistent with the fact that the staggered ordered state has
significantly lower energy than the random flux state at the flux density
$n_{\rm F}=1/3$~\cite{Koga2021Majorana}. The clusters having relatively
large critical flux field, i.e., 12b, 20b, and 24c, might be related to
a random flux state.

Furthermore, some clusters having non-flux-free ground state are shown
in Fig.~\ref{fig:cluster_SM}(b) as additional information. For the
clusters 12b, 16b, 16c, and 20c, the flux-free state and full-flux state
are (nearly) degenerate in the ground state. For the cluster 14b, the ground
state contains five $\langle W_p \rangle=1$ and two $\langle W_p \rangle=-1$.
For the cluster 8a, the ground state is full-flux state. These results might
be useful to consider the finite-size effects in the Majorana basis for
future studies.

\subsection{Repulsive interaction between fluxes}
\label{sec:app_b}
\begin{table}[tbh]
	\caption{Critical flux field $h_{\rm f,c}$(NN) and $h_{\rm f,c}$(NNN) for
		the clusters denoted by 24b, 24c, and 18a. The effective repulsive
		interaction is estimated as
		$V_{\rm eff}=h_{\rm f,c}$(NN)-$h_{\rm f,c}$(NNN).
	}
	\begin{center}
		\begin{tabular}{wc{1cm}wc{2.3cm}wc{2.3cm}wc{2.3cm}}
			\hline
			\hline
			cluster & $h_{\rm f,c}$(NN) & $h_{\rm f,c}$(NNN) & $V_{\rm eff}$ \\
			\hline
			24b    &  0.0404969256      &  0.0337186000      &  0.006778326 \\
			24c    &  0.0471150167      &  0.0401504917      &  0.006964525 \\
			18a    &  0.0338108475      &  0.0271080335      &  0.006702814 \\
			\hline
			\hline
		\end{tabular}
	\end{center}
	\label{tab:Veff}
\end{table}

If a flux-ordered state such as staggered order is more stable than random
flux state, there must be a kind of repulsive interaction between fluxes.
We now estimate the effective value of the repulsive interaction by applying
flux field $h_{\rm f}W_p$ to two plaquettes on the clusters exhibiting excited
flux states with two nonneighboring fluxes. Such clusters are 24b, 24c, and
18a in Fig.~\ref{fig:cluster_SM}(a). First, we obtain two critical flux fields:
(i) $h_{\rm f,c}$(NN) with flux field applied to two neighboring plaquettes
and (ii) $h_{\rm f,c}$(NNN) with flux field applied to two
next-nearest-neighbor plaquettes. Since there is a repulsion between fluxes,
$h_{\rm f}$(NN) is larger than $h_{\rm f}$(NNN). The obtained values are
shown in Table~\ref{tab:Veff}. Thus, we may estimate the effective value of
the repulsive interaction $V_{\rm eff}$ from the difference between
$h_{\rm f,c}$(NN) and $h_{\rm f,c}$(NNN). The obtained value of $V_{\rm eff}$
are very close for three clusters, i.e., $V_{\rm eff}\approx0.007$ in unit of
$|2K|$, although a finite-size scaling analysis is required to further confirm
this value.

\subsection{spin-spin correlations in the non-Kitaev QSL phase}
\label{sec:app_c}
As shown in FIG.~8, the region of
$-0.45 \lesssim J_{2,3} \lesssim 0.55$ for the disordered system is
characterized as a QSL based on the structureless static structure factor.
Part of the QSL phase, i.e., at $0.35 \lesssim J_{2,3} \lesssim 0.55$,
is recognized as non-Kitaev one due to the reduced $\langle W_p \rangle$
value. Here, we look at the spin-spin correlation functions
$\langle {\bf S}_i \cdot {\bf S}_j \rangle$
to further support this discrimination of Kitaev and non-Kitaev QSLs.
In a pure Kitaev model, only the NN correlations are finite and 
longer-range ones are zero; that is faithfully reproduced by the 24-site
calculations. Even away from the pure Kitaev limit, the spin-spin correlation
decays rapidly with distance if the system is in the Kitaev QSL state.

\begin{figure}[tbh]
	\includegraphics[width=\columnwidth]{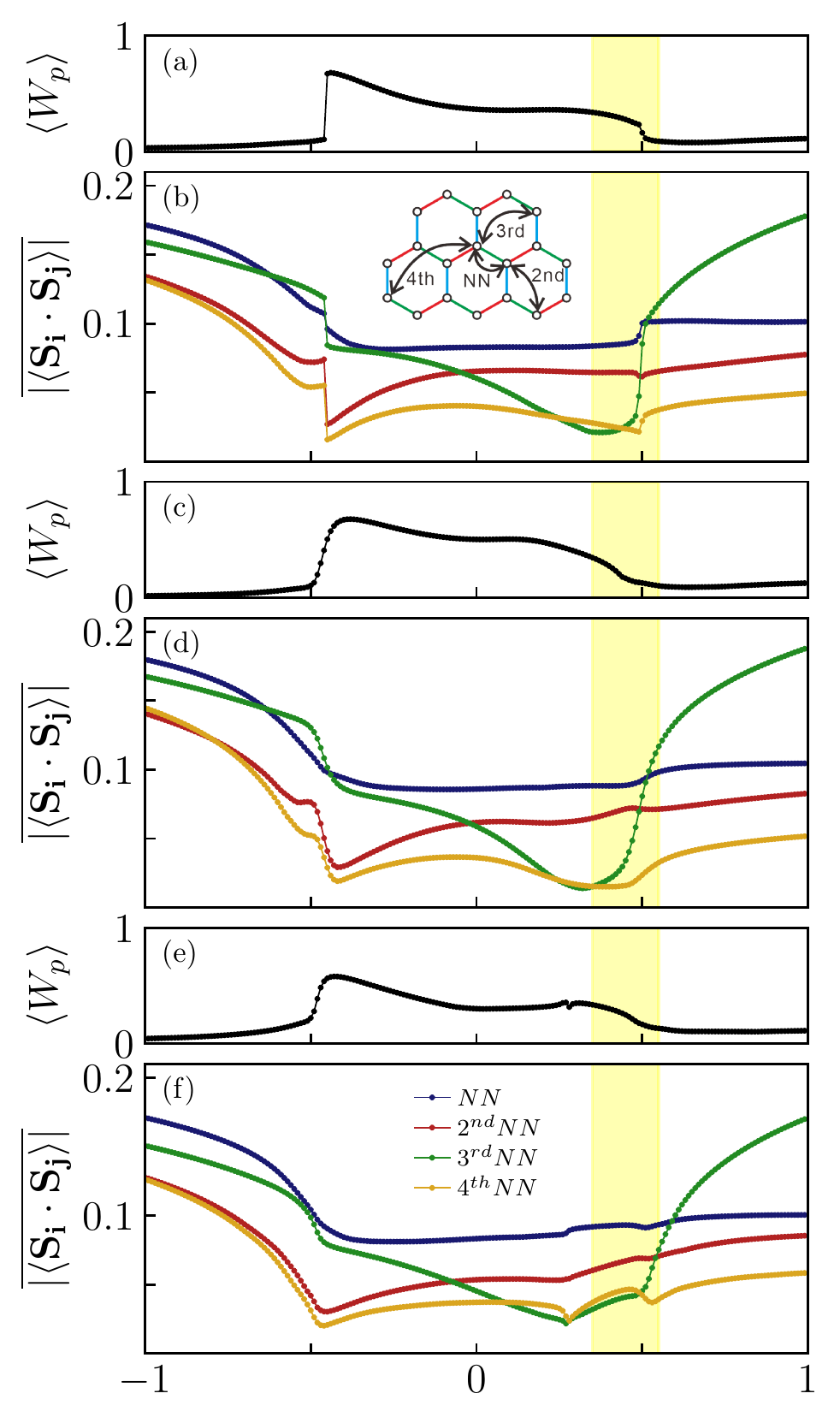}
	\caption{
		(a,c,e) Expectation value of flux operator and (b,d,f) spin-spin correlation
		functions for nearest-neighbor, 2nd-neighbor, 3rd-neighbor, and
		4th-neighbor bonds as a function of $J_{2,3}$ for three disorder
		samples. The inset in (b) shows the spin pairs considered and the yellow highlighted area shows the non-Kitaev spin liquid.
	}
	\label{fig:spincorr_SM}
\end{figure}

In Fig.~\ref{fig:spincorr_SM} the absolute values of spin-spin correlation
functions for nearest-neighbor, 2nd-neighbor, 3rd-neighbor, and 4th-neighbor
bonds are plotted as a function of $J_{2,3}$ for typical three
disorder samples. Since the system is disordered, the correlations are
averaged over all possible bonds. In the Kitaev QSL region,
we can see a rapid decay of
$\overline{|\langle {\bf S}_i \cdot {\bf S}_j \rangle|}$. Whereas in the
non-Kitaev QSL region, we find an enhancement of the 4th-neighbor correlation,
and it becomes larger than that for the 3rd-neighbor bond. This is clearly
in contrast to the Kitaev QSL feature. Interestingly, this reverse of
3rd-neighbor and 4th-neighbor correlations in the non-Kitaev QSL region
is confirmed in all of our 30 disorder samples.

\end{document}